\documentclass[twocolumn]{article}  

\usepackage{graphicx}
\usepackage{txfonts}

\def\Harpoons{\mathop{\rightleftharpoons}\limits}
\def\fs{\hbox{$.\!\!^{\rm s}$}}
\def\degr{\hbox{$^\circ$}}
\def\arcmin{\hbox{$^\prime$}}

\begin{document}

   \title{Detection of 6 K gas in Ophiuchus D}

   \author{J. Harju$^1$
          \and
          M. Juvela$^1$ 
          \and 
          S. Schlemmer$^2$
         \and
         L.K. Haikala$^1$ 
         \and 
         K. Lehtinen$^1$
         \and
         K. Mattila$^1$
\footnote{This publication is based on data acquired with the 
          Atacama Pathfinder Experiment (APEX). APEX is a collaboration
          between the Max-Planck-Institut f\"ur Radioastronomie, the
          European Southern Observatory, and the Onsala Space
          Observatory.}
          }


   \date{$^1$Observatory, P.O. Box 14, 
             FI-00014 University of Helsinki, Finland\\
              email: {jorma.harju@helsinki.fi} \\ 
         $^2$I. Physikalisches Institut,
              Universit\"at zu K\"oln, 
              Z\"ulpicher Stra{\ss}e 77, 
              D-50937 K\"oln, Germany}


   \maketitle

 
  \begin{abstract}

Cold cores in interstellar molecular clouds represent the very first
phase in star formation. The physical conditions of these objects are
studied in order to understand how molecular clouds evolve and how
stellar masses are determined.  The purpose of this study is to probe
conditions in the dense, starless clump Ophichus D (Oph D).  The
ground-state ($1_{10}\rightarrow1_{11}$) rotational transition of 
{\sl ortho}-H$_2$D$^+$ was observed with APEX towards the density peak of
Oph D.  The width of the H$_2$D$^+$ line indicates that the kinetic
temperature in the core is about 6 K. So far, this is the most direct
evidence of such cold gas in molecular clouds.  The observed
H$_2$D$^+$ spectrum can be reproduced with a hydrostatic model with
the temperature increasing from about 6 K in the centre to almost 10 K
at the surface. The model is unstable against any increase in the
external pressure, and the core is likely to form a low-mass star.
The results suggest that an equilibrium configuration is a feasible
intermediate stage of star formation even if the larger scale
structure of the cloud is thought to be determined by turbulent
fragmentation.  In comparison with the isothermal case, the inward
decrease in the temperature makes smaller, i.e. less massive, cores
susceptible to externally triggered collapse.
\end{abstract}


%

\section{Introduction}

Starless cores of molecular clouds are heated externally by the
interstellar radiation field (ISRF) and by cosmic rays.  Theoretical
estimates of the attenuation of the ISRF by dust suggest that the
temperature decreases to about 5-6 K in the centres of dense starless
cores (Evans et al. 2001; Zucconi et al. 2001; Stamatellos \&
Whitworth 2003).  Observational evidence of low temperatures is still
scarce.

Direct measurement of the gas properties, such as the kinetic
temperature, is possible using spectral lines. Spectral line
observations of very dense, starless cores are, however, hampered by
the fact that common tracer molecules like CO and CS freeze there onto
dust grains (Willacy et al. 1998; Caselli et al. 1999). The nitrogenous
compounds NH$_3$ and N$_2$H$^+$ seem to survive longer than most other
species (Tafalla et al. 2006). This is fortunate because the rotational
temperature of NH$_3$ is considered to be a good measure of the
kinetic temperature. Recently NH$_3$ excitation was used to derive a
very low gas temperature, $\sim 5.5$ K, in the centre of the prestellar
core L1544 in Taurus (Crapsi et al. 2007). 

At very high densities ($n \sim 10^6-10^7$ cm$^{-3}$) and low
temperatures ($T<10$ K) also NH$_3$ and N$_2$H$^+$ are likely to
freeze out, and the gallery of useful molecular tracers becomes very
limited (Walmsley et al. 2004). In these circumstances the abundances of
H$_3^+$, and its energetically favoured deuterated forms, H$_2$D$^+$,
D$_2$H$^+$, and D$_3^+$, are predicted to increase strongly. The
asymmetric isotopologues H$_2$D$^+$ and D$_2$H$^+$ have permanent
electric dipole moments, so can be used as probes of these
otherwise hidden regions (Caselli et al. 2003; Vastel et al. 2004; 
van der Tak et al. 2005).

Here we use the singly deuterated trihydrogen ion, H$_2$D$^+$, to show
that the gas kinetic temperature is about 6 K in the very dense core of
the starless clump Oph D (L1696A), which lies in the
nearby $\rho$ Ophiuchii molecular cloud.  The position observed is
located near the southern tip of the elongated clump. It corresponds
to the thermal dust emission maximum at $\lambda = 850 \mu$m on the
SCUBA map of Kirk et al. (2005), and the dust {\sl absorption} maximum on
the ISOCAM $7 \mu$m map of Bacmann et al. (2000).  This position
coincides with the N$_2$D$^+$ peak in the survey of Crapsi et al. 
(2005; see their Fig.~10), and represents the density and column density
maximum of the clump. The 3D structure of the whole clump has been 
recently modelled by Steinacker et al. (2005). 

\section{Observations and data reduction}

\begin{figure*}[htb]
\centering
\unitlength=1mm
\begin{picture}(160,72)(0,0)
\put(-5,0){
\begin{picture}(0,0) \includegraphics{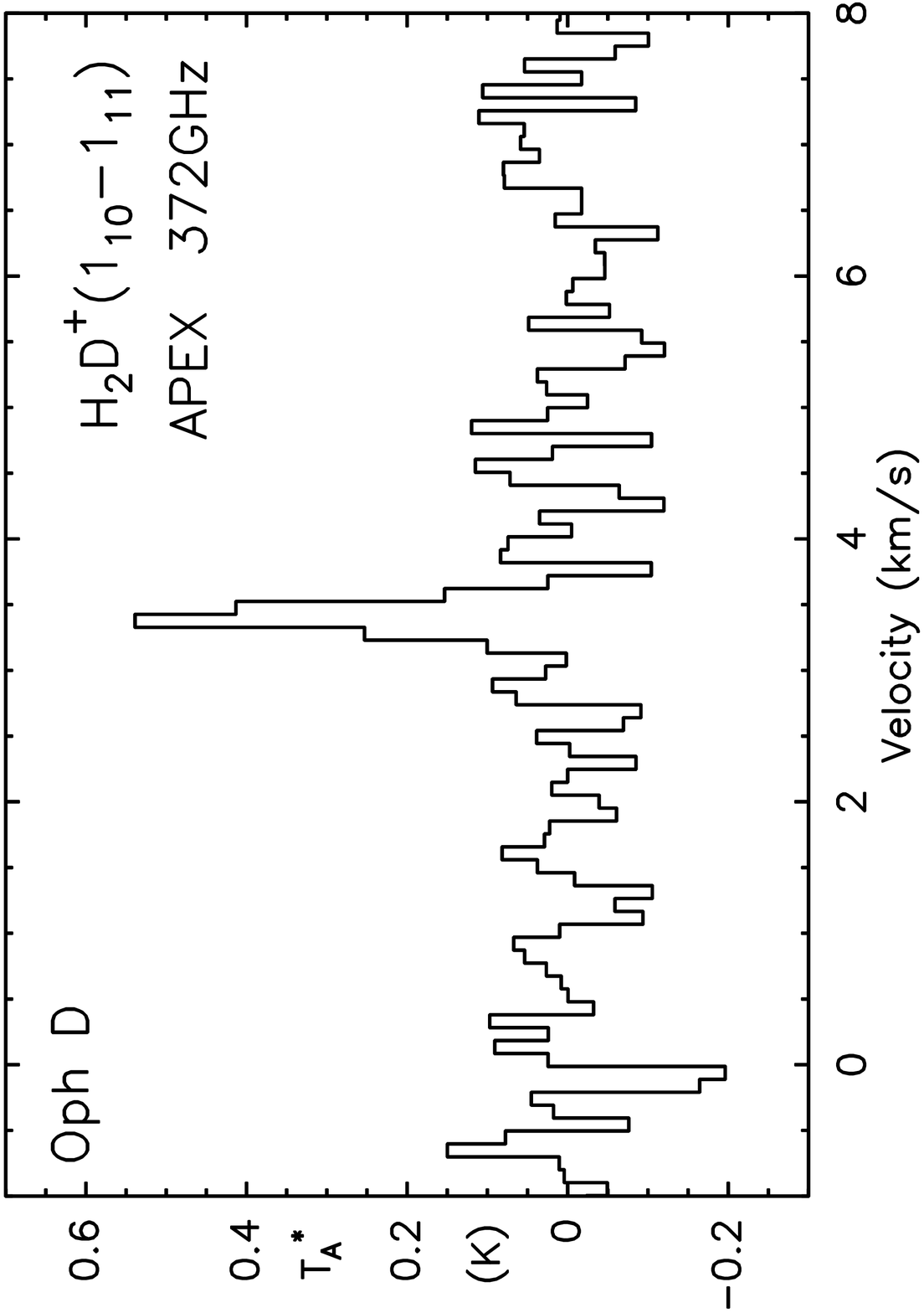} \end{picture}}
\put(98,0){\begin{picture}(0,0) \includegraphics{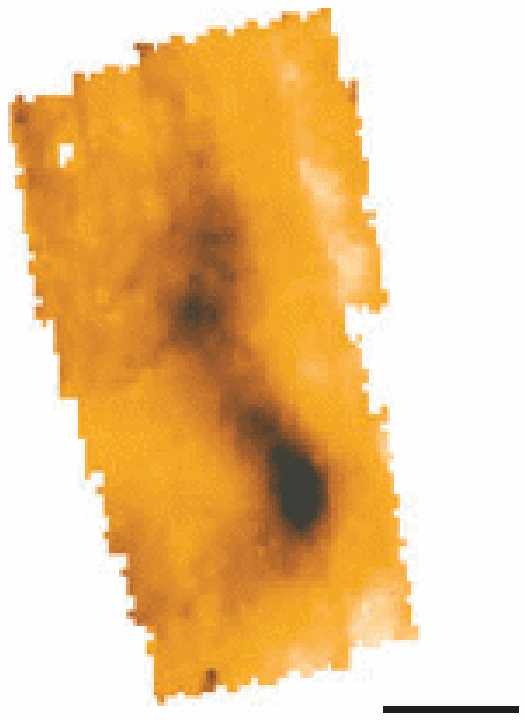} \end{picture}}
\put(85,25){\vector(1,0){40}}
\put(147,7){\makebox(0,0){$1^\prime$}}
\end{picture}
\caption{{\bf Left:} The H$_2$D$^+$ spectrum towards Oph D. 
{\bf Right:} ISOCAM 7$\mu$m mid-infrared absorption image of the
Oph D clump obtained by Bacmann et al. (2000; the image is
reproduced from Bergin \& Tafalla 2007). The scale is indicated by a
$1^\prime$ bar.}
\label{figure:h2d+spectrum}
\end{figure*}

The observations were made with APEX on May 15, 2006. The telescope is
described by G\"usten et al. (2006).  The 372421.364 MHz line of
{\sl ortho}-H$_2$D$^+$ was observed in the upper sideband of the
APEX-2A SIS DSB receiver. The HPBW of the antenna is $17^{\prime\prime}$ at
this frequency.  The backend was the Max-Planck-Institut f\"ur
Radioastronomie Fourier transform spectrometer (FFTS).  The 1 GHz band
of the FFTS was divided into 16384 channels resulting in a channel
width of 61 kHz which corresponds to $\sim$ 50 ms$^{-1}$ at the
observed frequency. The effective resolution of the spectrometer
corresponds to 80 ms$^{-1}$.  The observations were performed in the
position-switching mode. The coordinates of the ON-position are
R.A.$16^{\rm h}28^{\rm m}28\fs9$, Dec.$-24\degr19\arcmin09^{\prime\prime}$
(J2000). The OFF-position was selected $-10\arcmin$ south of
ON-position in Dec.  The spectrum shown in Fig.~1 has been obtained by
adding 60 ON-OFF measurements with a duration of 20 seconds per
phase. The observed position is indicated on the $7\mu$m ISOCAM
absorption map of Bacmann et al. (2000; reproduced from
Bergin \& Tafalla 2007).  The observing conditions were good (PVW 0.34 mm,
zenith opacity 0.33 at 372 GHz).  Oph D was in the elevation
range $60\degr-70\degr$, and the DSB system temperature was about
210~K during these measurements.

The observed spectra were reduced in a standard way using the
GILDAS software package\footnote{Grenoble Image and Line Data Analysis
Software package has been developed by IRAM-Grenoble, see {\tt
http://www.iram.fr/IRAMFR/GILDAS}}.  The occasional low-frequency
ripple in the individual raw spectra was first fit with a sinusoidal
baseline, after which possible higher frequency ripple was masked out in
Fourier space. After averaging the 60 individual spectra, a first-order 
baseline was subtracted.

\section{Results}

The detected H$_2$D$^+$ line has a single, narrow component.  A
Gaussian fit to the spectrum gives the following line parameters:
$T_{\rm A}^*=0.60\pm0.09$ K, $v_{\rm LSR}=3.40\pm0.01$ kms$^{-1}$, and
$\Delta v({\rm FWHM})=0.26\pm0.03$ kms$^{-1}$.  The molecular mass of
H$_2$D$^+$ is only 4 a.m.u. The observed linewidth, if assumed to be
caused only by thermal broadening, indicates a kinetic temperature of
$T_{\rm kin} = 6.0\pm1.4$ K. This is an upper limit as the estimate
neglects the possible non-thermal broadening and the slight
instrumental broadening, the contribution of which is about 0.01
kms$^{-1}$.

\section{Discussion}

The temperature obtained towards Oph D is similar to the
estimate for the nucleus of L1544 mentioned above
(Crapsi et al. 2007). The present determination is more direct than
that in L1544 because it does not involve modelling. These two results
using very different methods consolidate evidence of very
low temperatures inside starless dense cores.

The H$_2$D$^+$ position coincides with the N$_2$D$^+$ maximum of the
core with an exceptionally high N$_2$D$^+$/N$_2$H$^+$ column density
ratio (Crapsi et al. 2005). The enhancement of N$_2$D$^+$ is
caused by the precursor molecule N$_2$ reacting with increasingly
abundant H$_2$D$^+$, D$_2$H$^+$ or D$_3^+$ instead of the normal
isotopologue H$_3^+$. Carbon monoxide has been observed to be heavily
depleted in this core (Bacmann et al. 2002). The accumulated
observational results, i.e. CO depletion, a large degree of deuterium
fractionation, the detection of H$_2$D$^+$, and a very low kinetic
temperature, all fit qualitatively into the current picture of
chemistry in dense starless cores (Walmsley et al. 2004).

The picture is not complete, however, until we manage to build 
a physical model for the object to explain the observed H$_2$D$^+$
profile.  The dense central core observed here apparently has very little
turbulence and looks roundish in the N$_2$D$^+$
maps (Crapsi et al. 2005). It may therefore be in near hydrostatic
equilibrium.  As the H$_2$D$^+$ line, which probably originates in the
core centre, indicates a very low temperature, it is reasonable to
assume a radial temperature gradient.  We adopt the modified
Bonnor-Ebert model (Evans et al. 2001; Zucconi et al. 2001), i.e., a
self-gravitating, hydrostatic core with a temperature gradient caused
by the attenuation of the ISRF.  As the condensation is part of a
larger cloud, we assume, somewhat arbitrarily, that the obscuration by
the surrounding envelope corresponds to a visual extinction of $A_{\rm
V}=10^{\rm m}$.  This choice is of little importance because the
obscuration by the core itself must be much higher to produce
a temperature close to 6 K in the centre.

\subsection{Core model}

The density and temperature distributions in the core were solved in
an iterative manner. The density distribution was first solved for an
isothermal Bonnor-Ebert sphere (Alves et al. 2002), and the radial
temperature gradient was calculated using the ISRF and dust opacity
models adopted from the literature (Black 1994; Ossenkopf \& Henning
1994).  This temperature distribution was used to derive a new density
profile by integrating the equation of hydrostatic equilibrium.

We assumed that the gas temperature, $T_{\rm kin}$, is equal to the
dust temperature, $T_{\rm dust}$.  This assumption is generally
believed to be valid at densities relevant to this study ($> 10^5$
cm$^{-3}$, Burke \& Hollenbach 1983). Recently, Bergin et al. (2006)
found evidence of different gas and dust temperatures in the centre of
the globule B68 with $n({\rm H_2}) \sim 3\,10^5$ cm$^{-3}$. We note,
however, that the nearly 10 times higher density in Oph D is likely to
result in a closer dust-gas thermal coupling as compared with B68.

All radiative transfer calculations were made using our Monte Carlo
code (Juvela 1997). It was found that central densities in excess of
$10^6$ cm$^{-3}$ are needed to make the temperature decrease close to
6 K.  As a test, we compared the temperature profile calculation with
the results of Stamatellos \& Whitworth (2003). Identical temperature
distributions were obtained using the same models for the core
stucture, ISRF spectrum, and the dust opacity.

\subsection{H$_2$D$^+$ excitation}

The structure and rotational spectrum of H$_2$D$^+$ are well known
(Miller et al. 1989).  Like molecular hydrogen, H$_2$, the molecule has
two nuclear spin states, {\sl ortho} (hereafter o-H$_2$D$^+$, the H
nuclei have parallel spins), and {\sl para} (p-H$_2$D$^+$, opposite
spins). While radiative transitions are only possible between
rotational levels of the same nuclear spin state, collisions with
H$_2$ can also result in ortho-para conversion (e.g.,
Gerlich et al. 2002).

The state-to-state coefficients for the H$_2$D$^+$ + H$_2$ system are
not available. In the recent study of H$_2$D$^+$ profiles towards
L1544 van der Tak et al. (2005) used scaled radiative rates adopted from
Black et al. (1990), where the downward collisional coefficients within
the ortho and para ladders were assumed to be proportional to the line
strengths of the corresponding radiative transitions, and a constant
value was used for downward inter-ladder transitions. The upward 
rates were evaluated by using the principle of detailed balancing.
 
A microcanonical statistical study of the reaction H$_3^+$ + H$_2$ has
been published recently (Park \& Light 2007).  This treatment is being
applied to the deuterated isotopologues, but the state-to-state
reaction probabilities for H$_2$D$^+$ + H$_2$ have not yet been
calculated (Hugo et al. 2007). In the present paper the collisional
coefficients were calculated using a modification of the
Oka \& Epp (2004) model. The coefficients are normalized to conform with
the available chemical data on the nuclear spin-changing reactions.
The assumptions and parameter values used in the calculation are
available in the appendix. The approximation corresponds to
the canonical approach discussed in Hugo et al. (2007).

The core model and the collisional coefficients were used to calculate
the populations of the rotational levels of H$_2$D$^+$, and
H$_2$D$^+(1_{10}\rightarrow1_{11})$ line profiles towards the core
centre. The calculated spectrum corresponds to the source brightness
temperature distribution convolved with the $17^{\prime\prime}$ APEX
beam.  The central density and the outer radius of the core and the
total (o+p) H$_2$D$^+$ abundance were varied, and the resulting line
profile was compared with the observed one. Our best-fit H$_2$D$^+$
spectrum and the corresponding core model are shown in Fig.~2. The
model spectrum was converted to the $T_{\rm A}^*$ scale by
multiplying it by the main-beam efficiency, $\eta_{\rm MB} = 0.73$.

\begin{figure}
\unitlength=1mm
\begin{picture}(80,150)
\put(-7,48){
\begin{picture}(0,0) \includegraphics{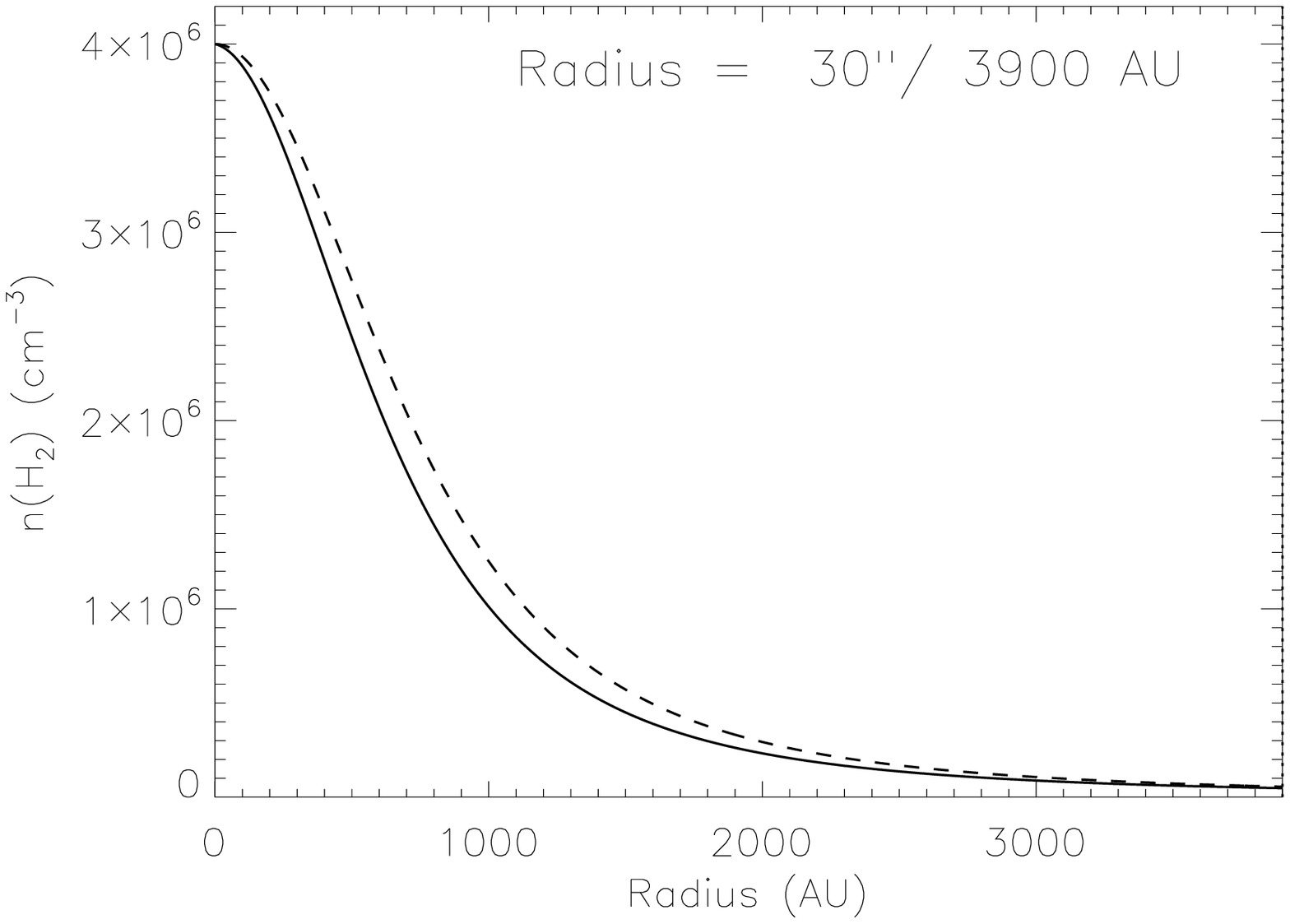} \end{picture}}
\put(-2,-3){\begin{picture}(0,0) \includegraphics{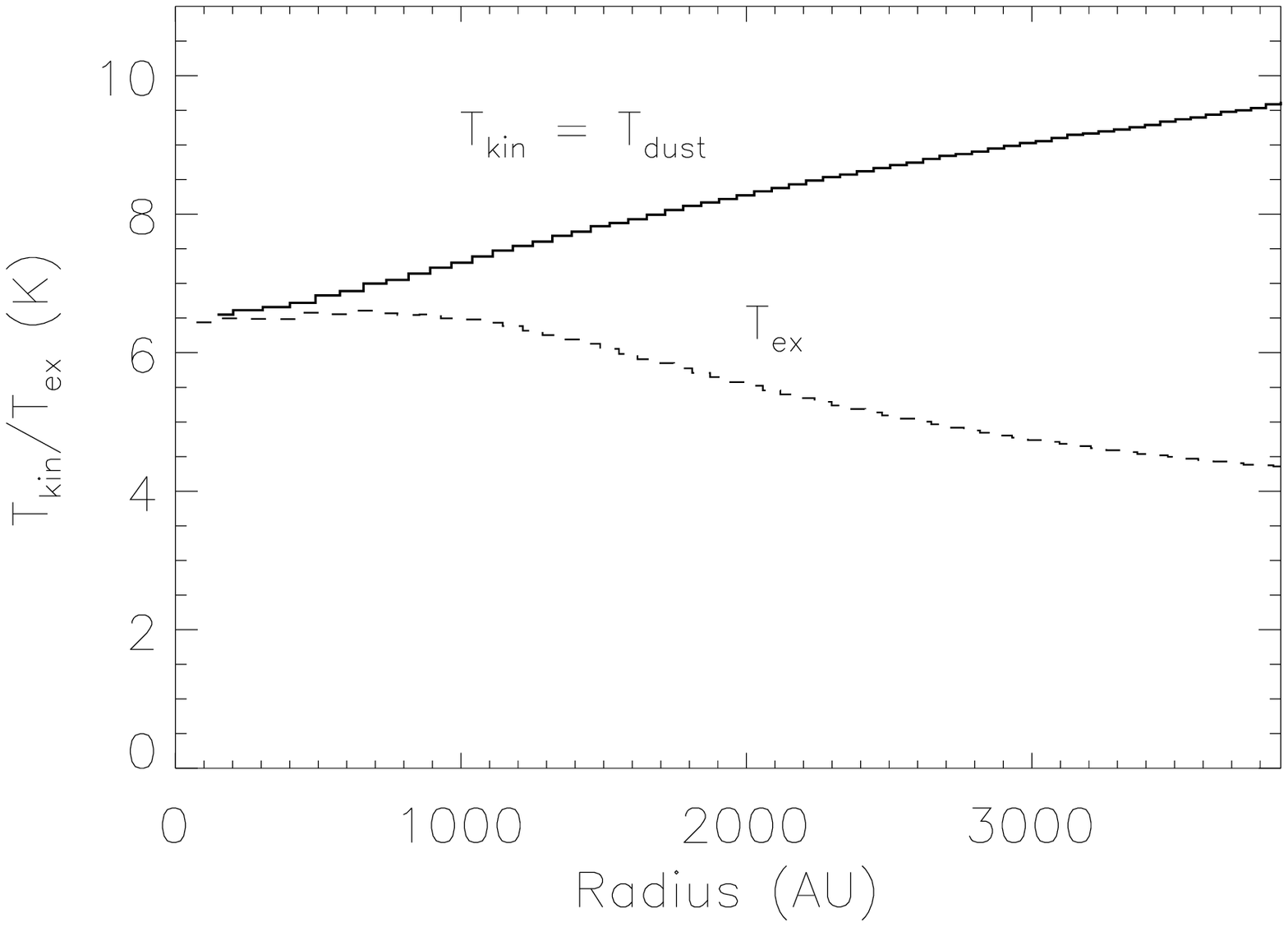} \end{picture}}
\put(-8,-54){\begin{picture}(0,0) \includegraphics{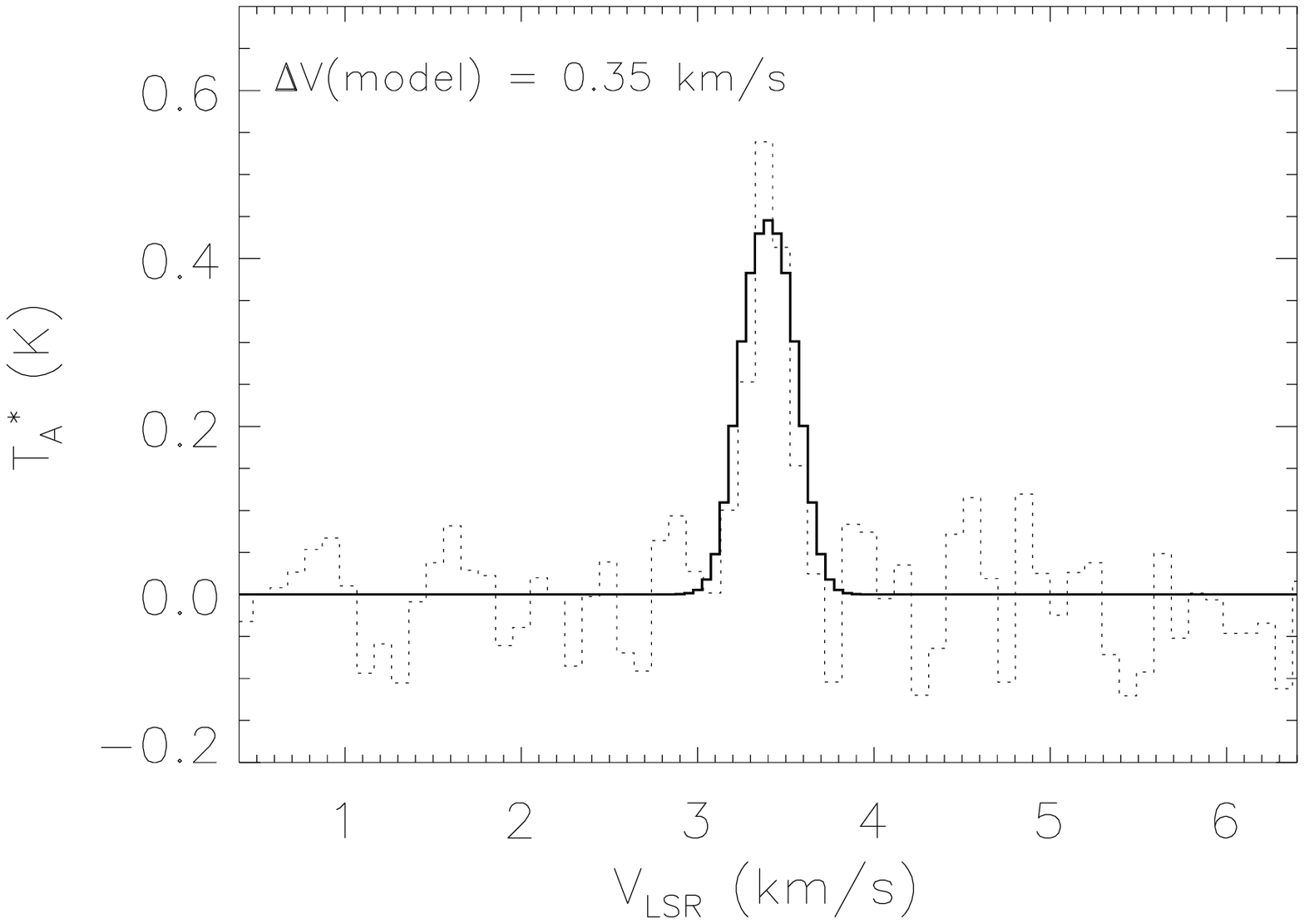} \end{picture}}
\end{picture}
\caption{The core model and the calculated H$_2$D$^+$ spectrum.  {\bf
Top:} Density as a function of radial distance from the core centre
according to our hydrostatic, non-isothermal model (continuous
line). For comparison the solution for an isothermal sphere with
$T_{\rm kin,iso} = <T_{\rm kin}> = 8.3$ K and the same peak density is
shown as a dashed line.  {\bf Middle:} The radial profiles of the
kinetic temperature, $T_{\rm kin}$ (assumed to be equal to $T_{\rm
dust}$), and the excitation temperature of the line, $T_{\rm ex}$. The
line is thermalized in the core centre, but becomes subthermally
excited towards the edge as the density decreases.  {\bf Bottom:}
H$_2$D$^+(1_{10}\rightarrow1_{11})$ spectrum (in the $T_{\rm A}^*$
scale) resulting from the Monte Carlo radiative transfer calculations
applied to the hydrostatic model presented in the two upper
panels. The observed spectrum is shown as a dotted line.}
\end{figure}

\subsection{Modelling results}

The line {\sl intensity} can be reproduced by a set of hydrostatic
models with the central density and the central temperature ranging
from $2\,10^6$ cm$^{-3}$ to $1\,10^{7}$ cm$^{-3}$, and from 6.8 K to
6.1 K, respectively.  The best-fit model has a central density of
$4\,10^6$ cm$^{-3}$. In this model the temperature decreases from 9.6
K at the surface to 6.4 K in the centre. The line {\sl width} of the
corresponding model spectrum is $0.35$ kms$^{-1}$. This and all model
spectra having similar peak intensity are broader than the observed
one. There are two reasons for this: 1) the kinetic temperature in the
centre of the model is slightly higher than 6 K, and 2) the peak
optical thickness of the line must be close to unity to give
sufficiently high intensity. Opacity broadening is responsible for
about one half of the `extra' linewidth ($\sim 0.1$ kms$^{-1}$) of the
best fit spectrum with $\tau_{\rm peak} \sim 1.1$. This $\tau$ value
results from a total H$_2$ column density of $1\, 10^{23}$ cm$^{-2}$
and a fractional H$_2$D$^+$ abundance of $5\,10^{-10}$.  The core mass
is nearly the same, 0.25 $M_\odot$, for all hydrostatic models quoted
above. The outer radius was assumed to be 3900 AU (apparent radius
30$^{\prime\prime}$). This is consistent with the published maps
(Kirk et al. 2005; Crapsi et al. 2005). We assume a distance
of 130 pc (Knude \&  H{\o}g 1998).

The core can be made cooler by increasing the central density further.
This would, however, decrease the line intensity because of a larger
optical thickness and beam-dilution effects as the brightness
distribution becomes more centrally peaked.


We also calculated temperature distributions and H$_2$D$^+$ spectra
for the density profiles derived previously for Oph D by
Motte et al. (1998) and Bacmann et al. (2000), who used millimetre dust
continuum emission and mid-infrared absorption, respectively. These
models assume a constant density within a radius of $20^{\prime\prime} -
25^{\prime\prime}$ from the core centre, and a power law in the outer envelope
up to $85^{\prime\prime}$.  For the Motte et al. (1998) model, with a central
density of $9\,10^5$ cm$^{-3}$, the calculated temperature goes down
to 6.1 K in the centre. The Bacmann et al. (2000) model with a lower
central density ($3\,10^5$ cm$^{-3}$) gives a minimum temperature of
7.4 K.  The model of Motte et al. produces similar H$_2$D$^+$ spectra
to the hydrostatic models quoted above. The existence of a large,
homogenous, and non-turbulent nucleus could possibly be justified by
invoking a stabilizing magnetic field.

In view of the moderate S/N of the observed H$_2$D$^+$ line and the
uncertainties related to the collisional coefficients, the hydrostatic
model gives a reasonably good fit to the observed spectrum. A high S/N
H$_2$D$^+$ spectrum towards this source, together with rigorously
derived collisional coefficients will show if the slightly `too bright
and too narrow' line remains a problem.  Assuming that the excitation
within the ortho and para ladders is thermal, the H$_2$D$^+$ spectrum
constrains the model parameters in the following way. 1) In order to
reproduce both the observed intensity and linewidth, the central
temperature must be about 6 K but not much less. The minimum
temperature sets a limit to the maximum density through the adopted
ISRF and dust opacity models. 2) The H$_2$D$^+$ line is preferably
optically thin, and in any case the peak optical thickness cannot be
much larger than unity. For high opacities, the line peak flattens and
the FWHM becomes clearly larger than observed.  This determines the
upper limit of the column density and the fractional abundance of the
molecule.  According to the adopted model for ortho-para transitions,
the o/p ratio of H$_2$D$^+$ increases as the temperature decreases,
and consequently, in the centre most H$_2$D$^+$ is in ortho
states. This effect facilitates the detection of H$_2$D$^+$ in cold
cores. Furthermore, the accuracy of the modelling results is improved
by nearly all H$_2$D$^+$ molecules being in the two states coupled by
the observed transition.

According to Steinacker  et al. (2005) the elongated structure of the
Oph D clump is likely to have formed as a result of turbulent
fragmentation.  They suggest that the southern condensation (i.e. the
core observed here) may be gravitationally bound and collapsing. The
present observation and modelling conform with the idea of a
self-gravitating core, but the very narrow line profile 
suggests that the collapse has temporarily halted.

Compared with an isothermal core at a kinetic temperature $T_{\rm
kin,iso}$, a core with inward decreasing temperature with the same
central density and the average temperature $<T_{\rm kin}> = T_{\rm
kin,iso}$ has a smaller critical radius (Alves et al. 2002) beyond
which the core becomes unstable against an increase of the external
pressure.  While for an isothermal sphere the critical dimensionless
radius parameter, $\xi$, is 6.45, the corresponding parameter for our
best-fit model is 6.20. This value corresponds to 1700 AU
($13^{\prime\prime}$), which is clearly smaller than the core radius
derived in previous studies.  Even though the effect might be
marginal, we note that the temperature gradient signifies that smaller
and less massive cores are susceptible to externally triggered
collapse.

\section{Conclusions}

The modelling results presented above depend on several unknown
factors, above all the collisional coefficients of H$_2$D$^+$ and the
possible local variations in the ISRF and the dust
properties. Nevertheless, the observed H$_2$D$^+$ spectrum can be
reproduced reasonably well by a self-consistent physical model.  The
strength of H$_2$D$^+$ is that it selectively traces the dense nuclei
of cold cores, and can therefore be used to study their internal
structure.

In Oph D we seem to be witnessing a quiescent phase of core
evolution.  The modelling results suggest, however, that the balance
can be easily disturbed and that the core will eventually collapse. The
possibility of a temporary equilibrium configuration is interesting as
it has been suggested that Oph D was formed by turbulent
fragmentation (Steinacker et al. 2005).  

The usefulness of H$_2$D$^+$ in studies of the early stages of star
formation has been predicted by astrochemists and modellers
(e.g. Bergin et al. 2002).  Emerging high-altitude radio telescopes,
most recently APEX, are providing empirical tests of these ideas.

\vspace{2mm}

{\sl Acknowledgements}~
We thank the APEX staff, especially Andreas Lundgren and
Felipe Mac-Auliffe, who performed the observations, and Per Bergman, who
carefully checked that the intensity calibration is correct.  The
Helsinki group acknowledges support from the Academy of Finland
through grants 1117206, 1210518, 115056, and 107701.

\vspace{2mm}

\large
{\bf References}
\normalsize


\noindent
Alves, J.F., Lada, C.J., Lada, E.A. 
2002, Nature 409, 159

\noindent
Bacmann, A., Andr{\'e}, P.,
Puget, J.-L., Abergel, A., Bontemps, S., Ward-Thompson, D. 2000, 
A\&A 361, 555

\noindent
Bacmann, A., Lefloch, B., 
Ceccarelli, C. et al. 2002, A\&A 389, L6 

\noindent
Bergin, E.A., Tafalla, M. 
2007, ARA\&A 45, 339

\noindent
Bergin, E.A., Alves, J., 
Huard, T., Lada, C.J. 2002, ApJ 570, L101

\noindent
Bergin, E.A., Maret, S., van
der Tak, F.F.S., et al. 2006, ApJ 645. 369

\noindent
Black, J.H., van Dishoek, E.F., 
Willner, S.P., Woods, R.C. 1990, ApJ 358, 459 
 
\noindent
Black, J.H. 1994, in 
{\sl The First Symposium of the Infrared Cirrus and Diffuse
Interstellar Clouds}, ASP Conf. Ser. 58, 355

\noindent
Burke, J.R., Hollenbach, D.J. 
1983, ApJ 265, 223

\noindent
Caselli, P., Walmsley, C. M., 
Tafalla, M., Dore, L., Myers, P.C. 1999, ApJ 523, L165

\noindent
Caselli, P., van der Tak,
F.F.S., Ceccarelli, C., Bacmann, A. 2003, A\&A 403, L37

\noindent
Crapsi, A., Caselli, P., Walmsley, 
C.M., et al. 2005, ApJ 619, 379

\noindent
Crapsi, A., Caselli, P., 
Walmsley, C.M., Tafalla, M. 2007, A\&A 470, 221

\noindent
Evans, N.J. {\sc II}, Rawlings,
J.M.C., Shirley, Y.L., Mundy, L.G. 2001, ApJ 557, 193

\noindent
Flower, D.R., 
Pineau des For{\^e}ts, G., Walmsley, C.M. 2006, A\&A 449, 621

\noindent
Gerlich, D., Herbst, E., Roueff,
E. 2002, Planet Space Sci. 50, 1275

\noindent
G\"usten, R., Nyman, L.-\AA., 
Schilke, P., Menten, K., Cesarsky C., Booth, R. 2006, A\&A 454, L13

\noindent
Hugo, E., Asvany, O., Harju, J., 
Schlemmer, S. 2007, in {\sl Molecules in Space and Laboratory}, 
meeting held in Paris, France, May 14-18, 2007, eds. 
J.L. Lemaire \& F. Combes, p. 119

\noindent
Juvela, M. 1997, 
A\&A 322, 943

\noindent
Kirk, J.M., Ward-Thompson, D., Andr{\'e},
P. 2005, MNRAS 360, 1506

\noindent
Knude, J. \& H{\o}g, E.\
1998, A\&A 338, 897

\noindent
Miller, S., Tennyson, J., 
Sutcliffe, B.T. 1989, Mol. Phys. 66, 429

\noindent
Motte, F., Andr{\'e}, P., Neri, R.
1998, A\&A 336, 150

\noindent
Oka, T., Epp, E. 2004, ApJ 613, 349

\noindent
Ossenkopf, V.H., 
Henning, T. 1994, A\&A 291, 943

\noindent
Park, K., Light, J.C. 2007,
J.Chem.Phys. 126, 044305-1--19

\noindent
Stamatellos,
D., Whitworth, A.P. 2003, A\&A 407, 941

\noindent
Steinacker, J., Bacmann, A.,
Henning, Th., Klessen, R., Stickel, M. 2005, A\&A 434, 167

\noindent
Tafalla, M.,
Santiago-Garc{\'\i}a, J., Myers, P.C., et al. 2006, A\&A 455, 577

\noindent
van der Tak, F.F.S., 
Caselli, P., Ceccarelli, C. 2005, A\&A 439, 195 

\noindent
Vastel, C., Phillips T.G., Yoshida,
H. 2004, ApJ 606, L127

\noindent
Walmsley, C.M., Flower,
D.R., Pineau des For{\^e}ts, G. 2004, A\&A 418, 1035

\noindent
Willacy, K., Langer, W. D.,
Velusamy, T. 1998, ApJ 507, L171

\noindent
Zucconi, A., Walmsley,
C.M., Galli, D. 2001, A\&A 376, 650


\appendix

\section{Approximation for the collisional rates of H$_2$D$^+$} 

In the calculation of the state-to-state rate coefficients, we have
adopted the approximation used by Oka \& Epp (2004) for H$_3^+$ + H$_2$
collisions.  This approximation is based on the principles of complete
randomness and a detailed balance of upward and downward transitions
between any two levels. This model does not consider any nuclear spin
restrictions.  It is known, however, that in cold gas the
conversion between the para and ortho states of H$_2$D$^+$ occurs
predominantly via collisions with o-H$_2$. The following reactive
transitions are possible (Gerlich et al. 2002; Walmsley et al. 2004):
\begin{eqnarray}
{\rm p-H_2D^+} + {\rm o-H_2}  & 
\Harpoons^{k_1^+}_{k_1^-}&  
{\rm o-H_2D^+} + {\rm p-H_2}~~~ \\
{\rm o-H_2D^+} + {\rm o-H_2}  & 
\Harpoons^{k_2+}_{k_2^-}& 
{\rm p-H_2D^+} + {\rm p-H_2}~~~ \;.
\end{eqnarray}

In molecular clouds the {\sl ortho/para} ratio of H$_2$, o/p-H$_2$, is
likely to be non-thermal (i.e., more o-H$_2$ than expected from
thermodynamic equilibrium where o/p-H$_2$ $\sim 9 e^{-170.4{\rm
K}/T}$) (Gerlich et al. 2002; Walmsley et al. 2004;
Flower et al. 2006). Consequently, reaction (A.1) pumps H$_2$D$^+$ from
para to ortho states with the help of the internal energy of o-H$_2$,
and this effect becomes more marked at low temperatures where the
endothermic ``backward'' reaction is inhibited.

Because of the small fraction of o-H$_2$, the total rates of the
nuclear spin changing reactions are likely to be much lower than those
derived from the Oka \& Epp approximation. We have therefore scaled the
Oka \& Epp coefficients to be consistent with the total rates of the
nuclear spin-changing reactions (A.1) and (A.2).  The assumptions used in
the calculation are summarised below. 

1) The total rate coefficient of rotationally inelastic and 
nuclear spin reactive H$_2$D$^+$ + H$_2$ collisions, $k_{\rm C}$, 
corresponds to that given by the Oka \& Epp (2004) approximation at the
same temperature. The total number of collisions per cm$^3$ and s is 
$k_{\rm C} n({\rm H_2}) n({\rm H_2D^+})$.  The coefficient $k_C$ can be 
expanded in terms of state-to-state coefficients, $C_{i,f}$, as  
\begin{equation}
k_{\rm C}  = \frac{\sum_i{g_{i}\,e^{-E_{i}/T}}\sum_f{C_{i,f}}}
                  {Z({\rm H_2D^+})} \;,
\end{equation}
where $i$ and $f$ refer to the rotational states of H$_2$D$^+$, and
$Z$ is the rotational partition function.  The Oka \& Epp 
coefficients $C_{i,f}$ are normalized so that $k_{\rm C}$ approaches
in warm gas the Langevin rate coefficient for ion-molecule collisions.

We note that our definition of the collisional rate, $k_{\rm C} n({\rm
H_2})$, includes collisions with both p-H$_2$ and o-H$_2$, and the
rates of both o-p and p-o conversions of H$_2$D$^+$.

2) When collisions dominate, the relative populations within the ortho
and para states approach those in thermal equilibrium, although 
 o/p\,H$_2$D$^+$ can be non-thermal. This is equivalent
to the assumption that the rotationally inelastic collisions, i.e.,
o-o and p-p transitions, are fast compared to nuclear spin changing
reactions. 

3) Reactions (A.1) and (A.2) describe nuclear spin-changing collisions
summed up over all states, and determine o/p\,H$_2$D$^+$.
The following conditions are obtained for the 
state-to-state coefficients $C_{{\rm p}i,{\rm o}f}$ and 
$C_{{\rm o}i,{\rm p}f}$ corresponding to para-ortho and ortho-para
transitions, respectively:
\begin{eqnarray}
\frac{\sum_i{g_{pi}\,e^{-E_{pi}/T}}\sum_f{C_{pi,of}}}{Z({\rm p-H_2D^+})} &=& 
{\rm o/p \, H_2} \, k_1^+ \, + \, k_2^- 
\\
\frac{\sum_i{g_{oi}\,e^{-E_{oi}/T}}\sum_f{C_{oi,pf}}}{Z({\rm o-H_2D^+})} &=&
{\rm o/p \, H_2} \, k_2^+ \, + \, k_1^- \; .
\end{eqnarray} 
The partition functions $Z({\rm p-H_2D^+})$ and $Z({\rm o-H_2D^+})$
have a common zero-energy level (the ground state of para-H$_2$D$^+$).
Equations (A.4) and (A.5) constrain the corresponding sums for the p-p and
o-o coefficients, $C_{{\rm p}i,{\rm p}f}$ and $C_{{\rm o}i,{\rm o}f}$,
through assumption 1).

We assume that o/p\,H$_2$D$^+$ can be obtained from the condition of 
chemical equilibrium: 
\begin{equation}
{\rm o/p\,H_2D^+} = \frac{{\rm o/p \, H_2} \, k_1^+ \, + \, k_2^-}
                         {{\rm o/p \, H_2} \, k_2^+ \, + \, k_1^-} \;.
\end{equation}
The coefficient $k_2^-$ is very small at low temperatures, and
Eq. (A.6) is essentially the same as Eq. (7) in Gerlich et al. 
(2002).  By substitution one finds that a thermal o/p\,H$_2$ ($\sim 9
e^{-170.4{\rm K}/T}$) makes also o/p\,H$_2$D$^+$ thermal ($\sim 9 \,
e^{-86.4 {\rm K}/T}$).
 
\vspace{1mm}

The collisional coefficients satisfy the principle of detailed balance, 
i.e., $n_i C_{if} = n_f C_{fi}$ when the relative populations are given
by the Boltzmann distribution. The normalization described above ensures 
that this is also true for ortho--para transitions 
even if o/p\,H$_2$D$^+$ is non-thermal. 

\vspace{1mm}

In practice, all state-to-state collisional rate coefficients were
initially calculated using the Oka \& Epp formula. The rate
coefficients {\sl between} the ortho and para states were then scaled
using (A.4) and (A.5).  Thereafter rate coefficients {\sl within}
ortho and para states were normalized so that assumption 1) is
satisfied.

We used the following values for the total rate coefficients in
reactions (A.1) and (A.2): $k_1^+ = 2\,10^{-9}$ cm$^3$s$^{-1}$, $k_1^-
= k_1^+ e^{-84{\rm K}/T}$, $k_2^+ = k_1^+/18$, and $k_2^- = k_2^+
e^{-257{\rm K}/T}$ (Walmsley et al. 2004, the $k_2$ coefficients are
slightly modified to account for the spin statistics). The o/p ratio
was assumed to be $10^{-4}$, which is believed to be characteristic
of a dense dark cloud at an advanced chemical stage
(Walmsley et al. 2004; Flower et al. 2006).

In very dense gas collisional transitions between the lowest
rotational states of both para- and ortho-H$_2$D$^+$ can compete with
radiative transitions. For example, the observed ground-state ortho
line ($1_{10} \rightarrow 1_{11}$) is thermalized in the core centre
(see Fig.~2, middle panel). 

The collisional coefficients derived here have a steeper temperature
dependence than those used by van der Tak (2005) which are
proportional to the square root of the kinetic temperature.  The
(downward) collisional coefficient for the $1_{10} \rightarrow 1_{11}$
transitions has, however,  a relatively smooth slope between $\sim 7$ 
and $\sim 9$ K, where the average value is $\sim 2\,10^{-10}$
cm$^3$s$^{-1}$. This number corresponds to the best-fit
value that van der Tak (2005) obtained by scaling the
Black et al. (1990) coefficient by a factor of $\sim 5$.

\end{document}